\documentstyle[epsfig, twocolumn]{esapub}

 \def\ave#1{\left\langle #1 \right\rangle}
 \def\s{{\rm (s)}}
 \def\ts{{\thinspace}}
 \def\eps{{\epsilon}}
 \def\vp{\varphi}
 \def\Rm#1{{\rm #1}}
 \def\SN{{\displaystyle{\rm S}}\over{\displaystyle {\rm N}}}
{\catcode`\@=11
\gdef\SchlangeUnter#1#2{\lower2pt\vbox{\baselineskip 0pt \lineskip0pt
  \ialign{$\m@th#1\hfil##\hfil$\crcr#2\crcr\sim\crcr}}}
}
\def\gtrsim{\mathrel{\mathpalette\SchlangeUnter>}}
\def\lesssim{\mathrel{\mathpalette\SchlangeUnter<}}

\begin{document}
\setlength{\parindent}{0pt}
\setlength{\parskip}{ 10pt plus 1pt minus 1pt}
\setlength{\hoffset}{-1.5truecm}
\setlength{\textwidth}{ 17.1truecm }
\setlength{\columnsep}{1truecm }
\setlength{\columnseprule}{0pt}
\setlength{\headheight}{12pt}
\setlength{\headsep}{20pt}
\pagestyle{esapubheadings}

\title{INVESTIGATING THE DARK MATTER DISTRIBUTION WITH NGST}
\author{{Peter Schneider$^1$ and Jean-Paul Kneib$^2$}\\
$^1${Max-Planck-Institut f\"ur Astrophysik, Postfach 1523,
D-85740 Garching, Germany} \\
$^2$ {Observatoire Midi-Pyr\'en\'ees, 14 Av. E. Belin, F-31400
Toulouse, France }}
\maketitle
\begin{abstract}
The developments summarized with the name ``weak gravitational
lensing'' have led to exciting possibilities to study the (statistical
properties of the) dark matter distribution in the Universe.
Concentrating on those aspects which require deep wide-field imaging
surveys, we describe the basic principles of weak lensing and discuss
its future applications in view of NGST (a) to determine the
statistical properties of the dark matter halos of individual
galaxies, (b) to determine the mass and the mass profile of very
low-mass clusters and groups at medium redshift and/or of more massive
clusters at very high redshift, and (c) to measure the power spectrum
of the matter distribution in the Universe in the non-linear regime,
thereby also obtaining a mass-selected sample of halos and providing a
means to investigate the scale- and redshift dependence of the bias
`factor'.\vspace {5pt} \\
\end{abstract}
\section{INTRODUCTION}

Ultra-deep observations with NGST will reveal the
distribution and the physical properties of very high-redshift
galaxies. Precursors of our present-day galaxies can be studied with
unprecedented accuracy, exploring regions of parameter space
(redshift, luminosity, size) which are currently white spots on our
`map of knowledge'. The remaining link between these new insights into
the evolution of galaxies and structure formation and cosmology is the
relation between optical/IR properties of galaxies and their
precursors and the mass of the halo in which they are
embedded. Bridging this gap between the luminous properties of matter
concentrations and modelling of the formation of structure is
essential for our understanding of the evolution of galaxies, groups,
and clusters, and that of the Universe as a whole.

Gravitational lensing is probably the only method which can provide
this missing link. Probing the tidal gravitational field of mass
concentrations (or, more generally, of the intervening mass
distribution) by the distortion of light bundles originating from
background galaxies, the mass of galaxies, groups, and clusters can be
determined individually, or statistically. The unique feature of
gravitational lensing is that the total matter distribution is probed,
independent of the physical nature or state of the matter, and
independent of symmetry assumptions.

The weak gravitational
lensing effect has been studied in great detail recently, and applied
successfully to several (classes of) systems; in particular, the
projected (dark) matter distribution of several relatively
low-redshift clusters of galaxies has been reconstructed, using both
parametric and parameter-free approaches, (e.g., Fahlman et al. 1994;
Smail et al. 1995; Squires et al. 1996a,b; Seitz et al. 1996, 1998;
Luppino \& Kaiser 1997a; Hoekstra et al. 1998), and the mass and
physical scale of galaxy halos has been constrained statistically,
employing the so-called galaxy-galaxy lensing effect (Brainerd et al.\
1996; Griffiths et al.\ 1997; Hudson et al.\
1998).  Furthermore, clusters of galaxies have been detected from
observing their tidal gravitational field through the image distortion
of faint background galaxies (e.g., Luppino \& Kaiser 1997b), thus
allowing to define a sample of 
clusters from their mass properties only, i.e., without referring to
their luminous properties (Schneider 1996).

Before continuing describing scientific applications of weak lensing,
we shall first outline the basic method.

\section{TIDAL DISTORTION OF GALAXY IMAGES}
A gravitational lens provides a map from the observer's sky to the
undistorted sky, caused by the gravitational field of the
deflector. The deflection angle is determined by the mass distribution
of the lens, and the mapping from the source to the observable images
depends in addition on the redshifts of the sources and the
lens. Light bundles are not only deflected as a whole, but due to the
tidal component of the gravitational potential, they are
distorted. The image of a circular source will, to first order, be an
ellipse. Thus, if all background galaxies were circular, the
observable ellipticity of galaxy images would provide a direct means
to measure the local tidal gravitational field of the lens. Since
galaxies are intrinsically not circular, this simple method is
impractical. However, since it can be assumed that the orientation of
galaxies is random, a tidal field would cause a preferred orientation
of the images, or, if ellipticity is considered as a two-dimensional
quantity, it would cause a net ellipticity.
The distortions are such that galaxy images are
preferentially aligned tangent to the `mass center', such as is
clearly visible for the giant luminous arcs.

Denote by $\eps^\s$ the complex intrinsic ellipticity of a galaxy
(defined in terms of its second order brightness moments), and let
$\eps$ be the corresponding complex ellipticity of the observed
image. The variable $\eps$ is defined such that for an image with
elliptical isophotes of axis ratio $r$ and orientation $\vp$ of the
major axis relative to a fixed reference direction, $\eps=(1-r)/(1+r)
\Rm e^{2{\rm i}\vp}$. The locally linearized gravitational lens
equation yields the transformation between $\eps^\s$ and $\eps$, which
for the case of weak distortions reads
$$
\eps=\eps^\s + \gamma\; ,
\eqno (1)
$$
and $\gamma$ is the shear or the tidal gravitational field of the lens
(for definition, see, e.g., Schneider, Ehlers \& Falco 1992)
defined in terms of the trace-free part of the Hessian of the
gravitational potential. The random orientation of intrinsic
ellipticities then implies that the expectation value of $\eps$ equals
$\gamma$, and the noise of this estimator is determined by the width
of the intrinsic ellipticity distribution: to measure a shear with
precision $\Delta\gamma$, one needs $N\sim
\sigma_\eps^2/(\Delta\gamma)^2$ galaxy images. Therefore, in order to
measure a weak shear signal, $N$ must be large, which can be achieved
by either increasing a local `smoothing scale' -- thereby erasing
small-scale structure information, or by increasing the number density
of galaxy images, that is, by taking very deep images.

In the past few years, observational progress was made possible by the
development of optical imaging instruments with extremely good image
quality and/or wide-angle capability. These instruments made it
possible to obtain images of thousands of faint galaxies (where faint
means $I\sim 24.5$, where the number density of galaxies is about 30
per square arcminute). Weak lensing observations with 8-meter class
telescopes will allow to obtain even deeper images; recent cluster
mass reconstructions obtained from Keck images used a number density
of about 60 galaxies/(arcmin)$^{2}$ (Clowe et al.\ 1998). However,
longer exposures do not guarantee a higher number density of useful
galaxy images, as fainter objects tend to become smaller: galaxies
much smaller than the extent of the seeing disk do not carry much
shape information and can therefore not be used. Seeing circularizes
images, and sophisticated methods to correct for this effect have been
developed (Bonnet \& Mellier 1995; Kaiser et al 1995; van
Waerbeke et al 1997). However, the factor by which the measured
ellipticity must be multiplied to get the true (before seeing)
ellipticity also multiplies the error of the measurement -- the
smaller the galaxy relative to the size of the seeing disk, the more
uncertain becomes the determination of 
the `true' ellipticity.

\section{WEAK LENSING WITH NGST}
Let us consider the specific example of a singular isothermal sphere
with l.o.s.\ velocity dispersion $\sigma_v$. For this specific mass
profile, an optimal statistics for the {\it detection} of this halo
can be defined, 
\begin{equation}
X\propto\sum_i {\epsilon_{{\rm t}i}\over \theta_i}\;,
\end{equation}
where $\theta_i$ is the separation of a background galaxy relative to
the center of the lens, and $\epsilon_{{\rm t}i}$ is the
tangential component of the image ellipticity (see, e.g., Schneider
1996). The sum extends over all background galaxies within an annulus
$\theta_1\le\theta\le\theta_2$, where the inner radis is constrained
by the ability to measure ellipticities reliably for images very close
to the foreground galaxy (or, in the case of clusters, where the
isothermal assumption breaks down due to flattening of the mass profile
near the center), and the outer radius is determined either by
considering neighboring foreground galaxies, or the decline of the
mass profile relative to an isothermal profile. The signal-to-noise
ratio of this statistics can be calculated to be
\begin{eqnarray}
{{\rm S}\over {\rm N}}&=& 12.7 \left( {n\over 30{\rm arcmin}^{-2}}
 \right)^{1/2}
\left( {\sigma_\epsilon \over 0.2 }\right) ^{-1} \\
&\times& 
\left(
 {\sigma_v\over 600{\rm km/s} }\right) ^{2}
\left( {\ln(\theta_2/\theta_1) \over \ln(10)}\right) ^{1/2}
\left\langle {D_{\rm ds}\over D_{\rm s}} \right\rangle \;,\nonumber
\end{eqnarray}
where $D_{\rm ds}$ and $D_{\rm s}$ are the angular-diameter distances
from the lens and the observer to the source, respectively; the
angular bracket denotes an average over the source population (the ratio
$D_{\rm ds}/ D_{\rm s}$ is set to zero for galaxies with smaller
redshift than the lens), and $\sigma_\eps$ is the dispersion of the
intrinsic ellipticity distribution. The larger S/N, the more details
of the mass profile can be inferred from weak lensing observations.

The foregoing estimate allows us to discuss the unique features of
NGST for weak lensing:

(1) Depth: the large aperture, combined with the largely reduced sky
background, allows the imaging of much fainter galaxies than currently
feasible. This will lead to a much larger number density $n$ of galaxy
images which can be used for these studies, and (most likely) to a
considerably higher mean redshift of the galaxies, increasing the
geometrical factor ${D_{\rm ds}/ D_{\rm s}}$.

(2) Angular resolution: Since galaxies tend to become smaller with
decreasing brightness, the number of galaxies that can be used for
weak lensing studies depends on the angular resolution; only galaxies
which are not significantly smaller than the size of the PSF can have
a reliable shape measurement. The unique angular resolution of NGST
will allow to measure the shapes of highest-redshift galaxies provided
they are larger in extent than $\sim 500 h^{-1}$ parsec, all of which
contribute to $n$. Hence, even if the 10-meter class ground-based
telescopes will carry out imaging surveys at significantly fainter
magnitudes than currently available, their use for the weak lensing
studies as outlined above will be limited, due to the small size of
the very faint galaxies compared to the ground-based PSF. In addition,
sampling of the PSF is essential for measuring accurate image
ellipticities.

(3) Wavelength range: Since NGST provides images in the NIR with
comparable or larger depth than even the HDF in the optical, it maps the
visual-to-red emission of high-redshift galaxies which is expected to
be considerably more regular than the rest-frame UV radiation. Since
the noise of mass measurements increases proportionally to the
dispersion $\sigma_\epsilon$ of the intrinsic ellipticities of the
background galaxies, mass estimates obtained from typical NGST
wavebands will be more accurate than from optical telescopes. In
addition, wavelength coverage from the visual to 5$\mu$ will allow
precise redshift estimates from multicolor photometry in which case
the ${D_{\rm ds}/ D_{\rm s}}$ ratio no longer is a statistical
quantity, but can be determined for each background galaxy separately,
which will allow the use of more sensitive estimators than (1). 

(4) Wide-angular field imaging: Nearly all applications of weak
lensing depend on solid angle $\omega_{\rm cam}$ covered per exposure:
for a given science goal, the total observing time scales like
$\omega_{\rm cam}^{-1}$. Compared to the WFC-part of the WFPC-2, the
currently planned 8K camera provides a gain of a factor $\sim 3.5$ in
$\omega_{\rm cam}$, and hardly any gain compared to the Advance
Camera.  However, {\it this is another one of the science cases where
switching to a larger-size CCD mosaic would immediately reduce the
observing time needed, or for given observing time, would enable more
ambitious science goals to be obtained.} E.g., switching to an 16K
mosaic, such as soon will become routinely available for optical
wavelength, would yield an advantage by a factor of 16 relative to the
WFC.

Quantitative predictions of the quantities relevant for NGST images
are highly model dependent and therefore uncertain. For example,
current number counts in the K-band extend to about $K\sim 24$,
whereas even a short exposure (20 minutes) with NGST will easily reach
$K\sim 27$ (for a S/N per object of 5). Hence, one has to extrapolate
at least 3 magnitudes into the dark. The limiting magnitude depends
also on the (unknown) size of these faint galaxies; in a few hours of
integration, $K=29$ (S/N=5) can be reached if the size of galaxies is
smaller than about 0.2 arcseconds. Roughly speaking, for a typical
integration time of one hour, one can reach a limiting magnitude of
about $K=28$ for extended objects with sufficient S/N, probably
yielding a number density of $\sim 2\times 10^6{\rm deg}^{-2}$, or a
factor of $\sim 20$ larger than those which are currently used for
weak lensing. In addition, the redshift distribution of these galaxies
will extend to much higher redshifts, increasing the geometrical
factor ${D_{\rm ds}/ D_{\rm s}}$; this effect is particularly
important for the high-redshift lenses.
Generally speaking, this implies
that {\it weak lensing studies can be extended with NGST to much less
massive and/or higher redshift matter concentrations than currently
possible.}

Next, three specific application shall be discussed in somewhat more
detail. 

\section{DETERMINING THE STATISTICAL PROPERTIES OF THE DARK MATTER
HALOS OF GALAXIES}
Individual galaxy halos are not massive enough to be `seen' with weak
lensing, but the lensing effect of many foreground galaxies can be
superposed. 

The basic principle here is to study a statistical alignment of the
images of background galaxies relative to their nearest neighboring
foreground galaxies. Assuming for a moment that the redshifts of all
galaxies are known, and that the massive properties of the galaxies
(such as a velocity dispersion and a truncation scale length) scale
with luminosity, the shear for each background galaxy can be
predicted, and a likelihood function in terms of the mass parameters
can be defined and maximized (Schneider \& Rix 1997).
If the redshifts are unknown, the
likelihood function has to be averaged over the magnitude-dependent
redshift probability distribution. 

In a first application using ground-based data, Brainerd et al (1996)
have shown that the rotational velocity of an $L_*$ field galaxy is about
220\ km/s, very much in accord with spectroscopically derived results,
but measured for higher-redshift ($\ave{z}\sim 0.3$) 
galaxies. Further results using the
Medium Deep Survey (Griffiths et al 1997) and the HDF (Dell'Antonio
\& Tyson 1996, Hudson et al
1998) have demonstrated the superiority of space-based imaging. In
these studies, the number of galaxy-galaxy pairs was not sufficient to
put upper bounds on the halo sizes of galaxies, but strict lower
bounds have been obtained.

NGST will allow to greatly extend these studies, in that the
achievable depth and superiour angular resolution and field-of-view
will permit to study higher-redshift galaxy halos, less massive halos,
and also avoid the use of excessive parametrization of the massive
properties of galaxies, but by investigating the halos of galaxies
with specific properties (such as photometric redshift estimates,
magnitude, colors, morphology).

Assuming $N_{\rm f}$ `identical' galaxy halos, the S/N for
their detection is $\propto N_{\rm f}^{1/2}$. To reach a given value
of S/N, one needs
\begin{eqnarray}
&N_{\rm f}=\displaystyle{350 \left( {{\rm S}/{\rm N}\over 10}\right) ^{2}
\left( {n\over 600{\rm arcmin}^{-2}}\right)^{-1}
\left( {\sigma_\epsilon \over 0.2 }\right) ^{2} } \\
&\times\displaystyle{
\left(  {\sigma_v\over 150{\rm km/s} }\right) ^{-4}
\left( {\ln(\theta_2/\theta_1) \over \ln(10)}\right) ^{-1}
\left( {\left\langle {D_{\rm ds}/ D_{\rm s}}
\right\rangle\over 0.15 } \right) ^{-2}}\nonumber
\end{eqnarray}
of these foreground galaxies ($D_{\rm ds}/ D_{\rm s}=0.15$ in an
EdS-Universe for $z_{\rm d}=3$ and $z_{\rm s}=5$). Note that a S/N of
10 means that $\sigma_v$ of these galaxies can be determined with a
5\% accuracy, or that additional parameters of the mass profile (such
as a truncation radius) can be determined. 

Investigating galaxies at $K\sim 24$ which will cover a broad redshift
interval $1\lesssim z\lesssim 5$ and which have a surface number
density of $\sim 
4\times 10^5 {\rm mag}^{-1}{\rm deg}^{-2}$, and splitting these
galaxies into bins of width $\Delta z=0.1$ and $\Delta K=0.1$ to
define `similar galaxies', we obtain about $10^3$ galaxies per bin and
per ${\rm deg}^{-2}$. Of course, different binning is also possible,
e.g., with larger $\Delta z$ (related to the accuracy of photometric
redshift estimates) and according to morphology or color. In any case,
the assumption of about 400 bins is physically meaningful.  Thus, for
the assumed fiducial parameters, $\sim 0.35{\rm deg}^2$ are needed to
obtain a S/N of ten in a typical bin. As was pointed out by Schneider
\& Rix (1997), and observationally verified by Hudson et al.\ (1998)
for the HDF, photometric redshifts will greatly increase the accuracy
of the mass parameters of galaxy halos; this is likely to be even more
relevant for high-redshift galaxies (where the redshift distributions
of galaxies with vastly different magnitudes will strongly overlap).

Galaxy-galaxy lensing will allow to study the evolution of halo masses
and size with redshift (at fixed luminosity), color, and morphology,
the verification and cosmological evolution of Tully-Fischer-like
scaling relations, and their dependence on environment (e.g., field
galaxies versus galaxies in pairs versus cluster/group galaxies -- see
Natarajan \& Kneib 1997; Geiger \& Schneider 1998a; Natarajan et al.\
1998), using maximum-likelihood techniques (Schneider \& Rix 1997;
Geiger \& Schneider 1998b). Whereas preliminary studies of
galaxy-galaxy lensing in clusters may indicate that the halo size of
galaxies in clusters is smaller than that for field galaxies, these
studies suffer from the very limited statistics currently
available. Cluster observations with the ACS will improve the
observational situation substantially, at least for medium-redshift
clusters, but the full power of this method can only be exploited by
the depth of NGST images which will allow an extension to
high-redshift clusters.  

In addition to weak lensing, given the depth of images routinely
obtained with NGST, 
these exposures will contain typically several strong lensing systems,
where a high-redshift faint galaxy will be mapped into two or more
images by a foreground galaxy, thus allowing an accurate estimate of
the mass of the latter. Note that several such candidate multiple
image systems have been found in the HDF (Hogg et al.\ 1996).
The probability of a given source at high
redshift to be multiply imaged depends on the evolution of the halo
abundance, but is typically a few $\times 10^{-4}$. With a number
density of $\sim 600 {\rm arcmin}^{-2}$, and a field of view of $\sim
16 {\rm arcmin}^2$, observing multiply imaged galaxies in virtually
every NGST image is certain. To identify them, the high angular
resolution of NGST to find morphologically similar galaxy images will
be of utmost importance, together with multi-color information. The
same observations required for the weak lensing programs mentioned
above will then be useful to constrain the halo abundance on mass
scales down to a few $\times 10^9 M_\odot$ as a function of redshift.

By combining these results on halo masses
with the optical/IR properties of galaxies, the missing link between
dark matter models and galaxy observations can be provided. 

\section{DETERMINING THE MASSES OF HIGH-REDSHIFT CLUSTERS/GROUPS}
The S/N estimate (2) shows that current ground-based weak lensing
studies can detect clusters with $\sigma_{\rm v}\gtrsim 600$\ km/s,
and determine the mass profile only for more massive clusters.
Changing to the NGST, a factor of $\sim 20$ in the galaxy
number density, plus the higher redshift of the background galaxies,
will change the `threshold' to $\sigma_{\rm v}\sim 250$\ km/s for
low-redshift clusters ($z\sim 0.3$), and/or allow to study the mass
profiles of much higher-redshift clusters.  

In order for clusters to be useful cosmological probes, their mass
properties need to be understood. Whereas current and future
wide-field imaging surveys will yield a large number of high-redshift
candidate clusters, the abundance of massive clusters needs to be
investigated by measuring their mass. Note that at higher redshift,
assumptions about dynamical and thermal equilibrium, or symmetry, are
probably less justified than at low redshift, making lensing an
essential tool for measuring the mass of high-redshift
clusters. Writing the estimate (2)  as appropriate for NGST, one finds
\begin{eqnarray}
&\SN&= 7.1 \left( {n\over 600{\rm arcmin}^{-2}}
 \right)^{1/2}
\left( {\sigma_\epsilon \over 0.2 }\right) ^{-1} \\
&&\times 
\left(
 {\sigma_v\over 250{\rm km/s} }\right) ^{2}
\left( {\ln(\theta_2/\theta_1) \over \ln(10)}\right) ^{1/2}
{\left\langle {D_{\rm ds}/ D_{\rm s}} \right\rangle\over 0.72}
\;,\nonumber
\end{eqnarray}
where the fiducial value for $D_{\rm ds}/ D_{\rm s}$ applies for
$z_{\rm d}=0.4 $, $z_{\rm s}=4  $ in an EdS Universe, or
\begin{eqnarray}
&\SN&= 6.1 \left( {n\over 600{\rm arcmin}^{-2}}
 \right)^{1/2}
\left( {\sigma_\epsilon \over 0.2 }\right) ^{-1} \\
&&\times  
\left(
 {\sigma_v\over 500{\rm km/s} }\right) ^{2}
\left( {\ln(\theta_2/\theta_1) \over \ln(10)}\right) ^{1/2}
{\left\langle {D_{\rm ds}/ D_{\rm s}} \right\rangle\over 0.155} \;,\nonumber
\end{eqnarray}
with the fiducial value of $D_{\rm ds}/ D_{\rm s}$ taken for $z_{\rm
d}=3 $, $z_{\rm s}=5 $. Hence one sees that the NGST will be able to
greatly extend the investigation of cluster masses and mass
distributions into regions of the mass-redshift space where currently
no reliable information on cluster masses is available. In particular,
the cosmic evolution of the morphology and the amount of substructure,
and its relation to the galaxy distribution (Kaiser \& Squires 1993,
Seitz \& Schneider 1995, 1996; Squires et al 1996a,b, 1997;
Seitz et al 1996, 1998; Hoekstra et al 1998) can be studied at high
redshift.

Different classes of cluster candidates can then be investigated:
those obtained from matched-filter analyses of optical images (e.g.,
Postman et al.\ 1996; Olsen et al.\ 1998), surface brightness
enhancement (Dalcanton et al.\ 1998), candidate clusters from SZ
surveys, as will be obtained, e.g., by the PLANCK Surveyor, 
extended sources from future X-ray missions etc. The various search
methods are likely to yield classes of clusters which differ in their
mass properties. Also, particularly interesting cases can be studied,
such as pairs of clusters, to search for connecting filaments, and
one might observe directly projected mass distributions which
characterize ongoing merging of cluster pairs. 

With background sources at typical redshifts of $z_{\rm s}\sim 5$, and
a sensitivity to very low-mass dark matter halos, the Universe becomes
`optically thick' to dark halos. Hence, along a given line-of-sight to
redshift 5, there will be more than one dark halo within a few
arcminutes which can be probed by (and which will affect) weak
lensing. Therefore, it is essential to `slice' the lenses into
redshift bins, or more practically, to use the sensitivity of the
lensing strength to the redshift of lens and source, expressed by  
$D_{\rm ds}/ D_{\rm s}$. This can be achieved by using redshift
information about the background galaxies, i.e., though relatively
accurate photometric redshifts. It should be noted that this argument
can be turned around: the dependence of the lensing strength on the
redshifts can be used to put constraints on the source redshift
distribution (Bartelmann \& Narayan 1995), and thus to test the
accuracy of photometric redshifts.  
Not much theoretical work has yet been done on such redshift-dependent
methods; future work, in combination with ray tracing through model
Universes obtained from N-body simulations (e.g., Wambsganss et al.\
1998; Jain et al.\ 1998) will clarify the power and limitations of
these applications.

Deep imaging of relatively low-redshift clusters will also be
extremely useful for strong-lensing applications. Given that a
few-orbit exposure with the HST-WFPC-2 reveals about ten strong
lensing candidate features (such as arclets; multiply-imaged galaxies)
in a strong cluster (e.g., A370 -- see Kneib et al.\ 1993; A2218,
Kneib et al.\ 1995, 1996; MS1512, Seitz et al.\ 1998), it is clear
that a moderately long exposure with NGST will probably detect of
order one hundred such strong lensing features; and probably also more
reliably than with WFPC-2, due to the better angular resolution which
is needed to compare candidate multiple images by their morphology. We
show in Fig.\ts 1 a first hint of what an NGST image of the cluster
A2218 might look like.  Also here, multi-color images will be
tremendously helpful. With these numerous strong lensing constraints
in a single cluster, the projected mass distribution of these clusters
can be determined with unprecedented spatial resolution, and even very
small subcomponents will be detectable.  In addition, each strong
lensing cluster will produce many highly magnified galaxy images, so
that the already truly impressive light-collecting power of NGST can
be coupled with these natural telescopes to map even fainter
galaxies. The `transition zone' between strong and weak lensing, in
which hundreds of arclets will be found, can be used to obtain
redshift estimates from the degree of their distortions, thus
providing an independent means to measure the redshifts of the
faintest galaxies observable with NGST (see, e.g., Kneib et al 1994;
Ebbels et al.\ 1996, 1998).

\begin{figure*}[!ht]
  \begin{center}
    \leavevmode
	\vspace{-2cm}
  \centerline{\epsfig{file=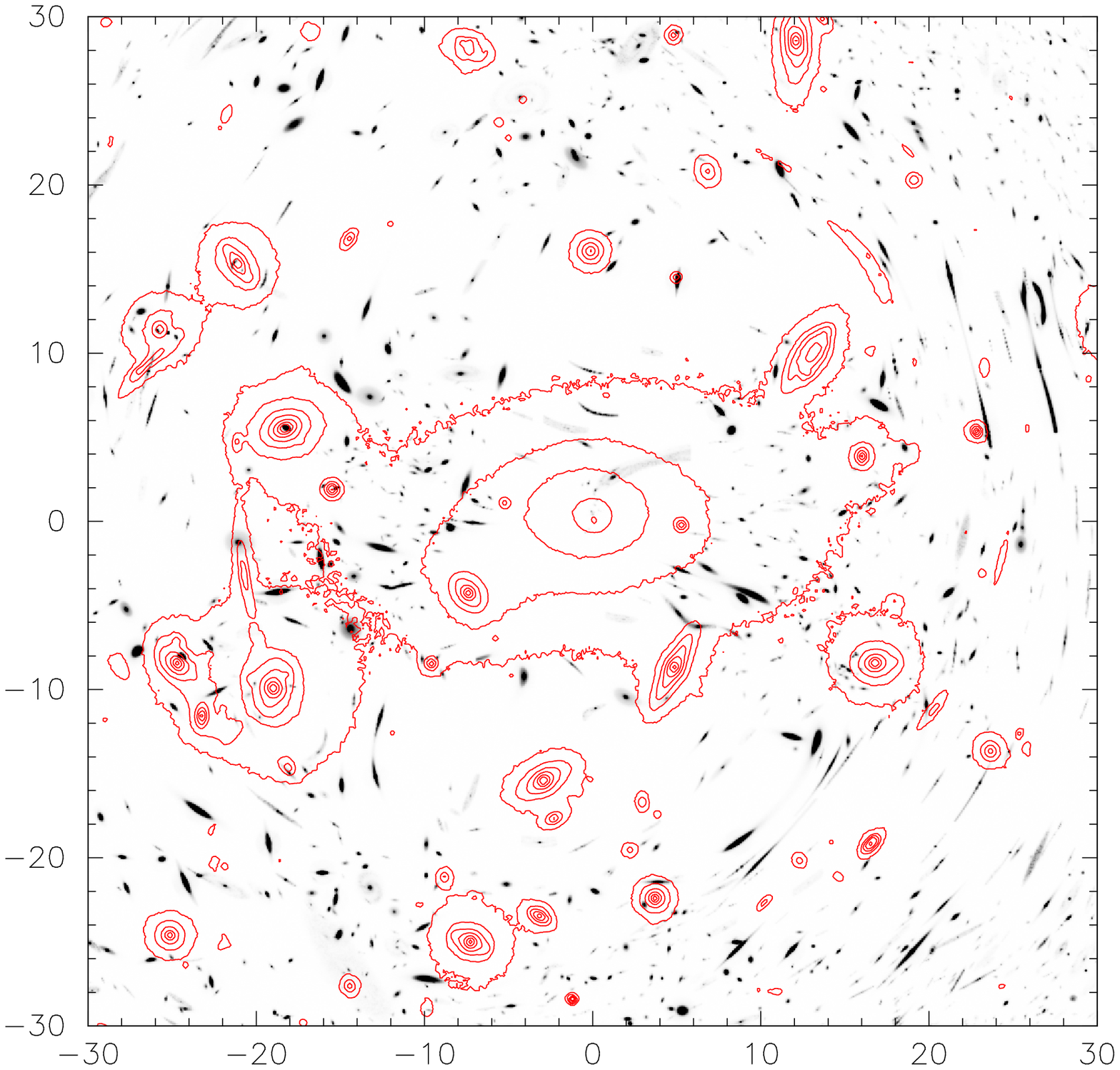,width=\textwidth}}
  \vspace{-0.7cm}
  \end{center}
  \caption{\em Simulated image of lensed features in the very central
  part of the massive cluster A2218. For these simulations, the mass
  profile of the cluster as constrained from HST observations and
  detailed modelling (Kneib et al 1996) has been used. The galaxy
  population model used in this simulation is somewhat simple (and
  likely to be highly oversimplied at such faint magnitudes).  The
  number density of (unlensed) sources was assumed to be $4\times 10^6
  {\rm deg}^{-2}$ down to K=29. To determine the physical quantities
  like absolute magnitude and intrinsic size, we applied the K+e
  correction from Bruzual and Charlot (1993) galaxy synthesis
  model. To match current observations, we also include a
  size-dependence with redshift. The redshift distribution assumed is
  broad and extend to redshift $z\sim10$ with a median value $z_{\rm
  med}\sim 3$.  The brighter objects (cluster galaxies and brightest
  arcs) seen by HST are displayed as contours, to make the faint
  galaxy images visible on this limited dynamic range reproduction. An
  enormous number of large arcs and arclets are seen; in particular,
  numerous radial arcs can be easily detected, which will allow to
  determine the `core size' of the cluster mass distribution. Due to
  the broad redshift distribution of the faint galaxies, arcs occur at
  quite a range of
  angular separations from the cluster center; this effect will become
  even stronger for higher-redshift clusters. It should be noted that
  this 1 arcminute field does not cover the second mass clump seen
  with HST; an NGST image will cover a much larger area, and more
  strong lensing features will be found which can then be combined
  with the weak lensing analysis of such a cluster.  For this
  simulation we have used a 0.06 arcsecond pixel size; the NGST
  sampling will be better by a factor of 2.
 }
\end{figure*}

The detailed mass models of clusters over a broad range of redshifts,
and its comparison with other observables (such as X-ray emission,
SZ-depletion, and galaxy distribution) will permit a thorough study of
the relation between the dark and the baryonic matter in
clusters. Questions like the validity of hydrostatic equilibrium of
intra-cluster gas, the gas-to-mass fraction as a function of cluster
mass and redshift, the possible presence of non-thermal pressure
etc. can be investigated in great detail.

Cluster with such a large number of strong lensing features also
provide a direct and purely geometrical handle on cosmological
parameters, again through the dependence of the lensing strength on
$D_{\rm ds}/D_{\rm s}$ (e.g., Link \& Pierce 1998).
Whereas this method could in principle also be
applied to current observations of multiple strong lensing features in
cluster, most of the strong lensing features are too faint to obtain a
reliable redshift, and in addition, one can also trade details of the
mass distribution for changes in $D_{\rm ds}/D_{\rm s}$. This will not
be the case if the number of strong lensing features becomes much
larger.

\section{MEASURING THE STATISTICAL PROPERTIES OF THE DARK MATTER 
DISTRIBUTION IN THE UNIVERSE}
The statistical properties of the
distortion field of the images of high-redshift galaxies directly
reflects the statistical properties of the intervening mass
distribution. For example, the two-point ellipticity statistics
(such as rms inside circular apertures, or correlation functions)
depend directly on the appropriately weighted and projected power
spectrum of the cosmic mass distribution (Blandford et al 1991; Kaiser
1992, 1998; Jain \& Seljak 1997; Schneider et al 1998b; Seljak 1997). A
preliminary detection of cosmic shear has been reported (Fort et al
1996; Schneider et al 1998a; Seitz 1998).

Cosmic shear is, besides CMB measurements, the cleanest probe of the
LSS, since it makes no assumption on the relation between mass and
light. Whereas the CMB can probe the LSS on comoving scales above
about 10 Mpc, cosmic shear is sensitive also to much smaller scales;
in particular, it can probe the non-linear evolution of the power
spectrum. A comparison with the linear power spectrum then yields
strong constraints on the evolution of the mass distribution, e.g.,
that of the power spectrum, at relatively late epochs. The
gravitational instability picture of structure formation can therefore
be tested with high accuracy. In addition, deep `blank-field' imaging
will also be used to define mass-selected samples of cluster, and in
particular might find group- and/or cluster-mass halos which are very
underluminous and therefore missed in standard cluster searches (`dark
clusters'). Such isolated halos form the highly non-Gaussian tail of
the LSS, which provides invaluable information for the distinction
between different cosmological model which may have the same projected
linear power spectrum, or the same two-point statistics (e.g., Jain et
al.\ 1998).

Ongoing and planned ground-based deep wide-field
imaging surveys will measure the power spectrum and the halo abundance
for typical redshifts of about 0.5; with NGST, this can be extended to
redshifts of order 2 or 3 which requires a much higher number density
of background galaxies and a considerably higher mean redshift for
them. By comparing the projected dark matter distribution,
statistically sliced into redshift bins according to the
(magnitude-dependent) redshift distribution of the background galaxies
(Seljak 1998), with the galaxy distribution at $z\lesssim 3$, the
redshift- and scale dependence of the bias `factor' can be studied in
great detail (Schneider 1998; van Waerbeke 1998). For example, to
obtain a statistically significant sample of $\sim 1000$ halos detected
from their mass properties only, a total area
of $\sim 10{\rm deg}^2$ mapped down to $K\sim 27$ would be sufficient.
This estimate depends strongly on the cosmological model (Kruse \&
Schneider 1998), and a
conservative number -- which applies to an EdS Universe -- has been
quoted. For an open model, the expected number of halos is about 3
times larger. 
As mentioned before, for such high redshifts the `dark halo confusion
limit' is reached (cf. the simulated shear maps by Jain et al 1998),
so photometric redshift information of the {\it background} galaxies
is essential to distinguish the tidal gravitational field of several
mass concentrations along the same line-of-sight, using the redshift
dependence of the lens effect.

Such a wide-field survey with NGST would also allow to directly
search for arcs. As was pointed out by Bartelmann et al (1998), the
statistics of arcs provides a very powerful handle on the cosmological
parameters, with expectations differing by a factor $\sim 100$ between
an EdS and an open Universe. They have shown that from the already
known number of arcs, a cluster-normalized EdS model is highly
disfavoured. This argument could be very much strengthened if a `field
survey' for arcs is untertaken. Since the majority of observed arcs
are very thin, partly unresolved even with WFPC-2, the angular
resolution provided by NGST will be invaluable for this search.

A first taste for the power of NGST to measure cosmic shear can be
obtained from the STIS instrument onboard HST: in a 20 minute exposure
taken with the CLEAR `filter', a galaxy number density of 50 -- 100
arcmin$^{-2}$ can be achieved; part of an ongoing parallel observing
program will be used to measure cosmic shear on the (small) scale of
STIS, 51\ arcseconds, and first results have been reported by Seitz
(1998). But NGST, with its much larger FOV and collecting area, and
smaller pixel size will be tremendously more powerful.

\section{DISCUSSION}
The previous examples should have illustrated the power of NGST for
weak lensing, i.e., to map the dark matter distribution at high
redshifts. The discussion here has been at most semi-quantitative, for at
least two reasons: First, NGST will probe the galaxy distribution at
luminosities and redshifts that are fully unexplored up to
now. Therefore, the statistical properties of the galaxy population
seen by NGST (number density, redshift distribution, size
distribution, etc.)  can be estimated only very roughly. Second, weak
lensing with NGST enters a regime which has not been considered
seriously up to now, even by daring theoreticians. Most probablematic
is the crowding along the lines-of-sight to redshifts $z_{\rm s}\sim
5$, mentioned before, and new statistical and theoretical tools need
to be developed and tested with numerical simulations. However, by
basing the discussion on the S/N estimate (2), the qualitative
estimates presented here should closely approximate the true power of
NGST for weak lensing. Furthermore, our estimates were based on the
assumption that K-band images will be used for weak lensing -- this is
not at all obvious. Certainly going to the NIR has the advantage of
probing the rest-frame optical wavelengths, but it may be that a
somewhat longer wavelength will turn out to be more efficient -- but
in the L or M band, even less is known about the galaxy population.

It should be stressed here that predictions for weak lensing
applications with NGST are extremely daring, and that the present
discussion provides a severe underestimate of these possibilities. To
wit, assume 10 years ago someone would have been asked to predict
the lensing applications of HST, or the VLT. At that time, arcs have
just been discovered, and Fort et al (1988) published the first
arclets. But none of the applications discussed above have been
seriously mentioned in the literature, with the first weak lensing
observation published in 1990 (Tyson et al 1990). In fact, only ten years
ago the dense population of background galaxies was pointed out
clearly by Tyson (1988). Given the strong rate of evolution in this
field, it is easy to predict that the range of weak lensing
applications of NGST are currently nearly unpredictable, in the sense
that they are likely to be much broader then what was presented here.

\section*{ACKNOWLEDGEMENTS}
We would like to thank Y.\ Mellier, R.\ Pello and S.\ Seitz for
numerous discussions. This work was supported by the
``Sonderforschungsbereich 375-95 f\"ur Astro--Teil\-chen\-phy\-sik"
der Deutschen For\-schungs\-ge\-mein\-schaft and the TMR Network
``Gravitational Lensing: New Constraints on Cosmology and the
Distribution of Dark Matter'' of the EC under contract
No. ERBFMRX-CT97-0172. PS would like to thank the Observatoire de
Midi-Pyren\`ees where most of this work has been done for its
hospitality and support.


\begin{thebibliography}{}

\bibitem{}{}{}
Bartelmann, M., Huss, A., Colberg, J.M.,
Jenkins, A. \& Pearce, F.R.\ 1998, MNRAS, in press (also astro-ph/9709229).

\bibitem{}{}{}
Bartelmann, M. \& Narayan, R.\ 1995, ApJ 451, 60.

\bibitem{}{}{}
Blandford, R.D., Saust, A.B., Brainerd, T.G. \& Villumsen, J.V.\ 1991,
MNRAS 251, 600.

\bibitem{}{}{}
Bonnet, H. \& Mellier, Y.\ 1995, A\&A 303, 331.

\bibitem{}{}{}
Bonnet, H., Mellier, Y. \& Fort, B.\ 1994, ApJ 427, L83.

\bibitem{}{}{}
Brainerd, T.G., Blandford, R.D. \& Smail, I.\ 1996 ApJ 466, 623.

\bibitem{}{}{}
Bruzual, G. \& Charlos, S.\ 1993, Apj 405, 538.

\bibitem{}{}{}
Clowe, D., Luppino, G.A., Kaiser, N., Henry, J.P. \& Gioia, I.\ 1998,
ApJ 497, L61.

\bibitem{}{}{}
Dalcanton, J., Spergel, D.N., Gunn, J.E., Schmidt, M. \& Schneider,
D.P.\ 1998, AJ 114, 2178.

\bibitem{}{}{}
Dell'Antonio, I. \& Tyson, J.A.\ 1996, ApJ 473, L17.

\bibitem{}{}{}
Ebbels, T.M.D.\ et al.\ 1996, MNRAS 281, L75.

\bibitem{}{}{}
Ebbels, T.M.D.\ et al.\ 1998, MNRAS 295, 75.

\bibitem{}{}{}
Fahlman, G., Kaiser, N., Squires, G. \& Woods, D.\ 1994, ApJ 437, 56.

\bibitem{}{}{}
Fort, B., Prieur, J.L., Mathez, G., Mellier, Y. \& Soucail, G.\ 1988,
A\&A 200, L17.

\bibitem{}{}{}
Fort, B., Mellier, Y., Dantel-Fort, M., Bonnet, H. \& Kneib, J.-P.\
1996, A\&A, 310, 705.

\bibitem{}{}{}
Geiger, B. \& Schneider, P.\ 1998a, MNRAS 295, 497.

\bibitem{}{}{}
Geiger, B. \& Schneider, P.\ 1998b, MNRAS, submitted.

\bibitem{}{}{}
Griffiths, R.E., Casertano, S., Im, M. \& Ratnatunga, K.U.\ 1996,
MNRAS 282, 1159.

\bibitem{}{}{}
Hoekstra, H., Franx, M., Kuijken, K. \& Squires, G.\ 1998, ApJ in
press, also astro-ph/9711096.

\bibitem{}{}{}
Hogg. D.W., Blandford, R.D., Kundic, T., Fassnacht, C.D. \& Malhotra,
S.\ 1996, ApJ 467, L73.

\bibitem{}{}{}
Hudson, M.J., Gwyn, S.D.J., Dahle, H. \& Kaiser, N.\ 1998, ApJ, in
press, also astro-ph/9711341.

\bibitem{}{}{}
Jain, B. \& Seljak, U.\ 1997, ApJ 484, 560.

\bibitem{}{}{}
Jain, B., Seljak, U. \& White, S.D.M.\ 1998, astro-ph/9804238.

\bibitem{}{}{}
Kaiser, N.\ 1992, ApJ 388, 272.

\bibitem{}{}{}
Kaiser, N.\ 1998, ApJ 498, 26.

\bibitem{}{}{}
Kaiser, N. \& Squires, G.\ 1993, ApJ 404, 441.

\bibitem{}{}{}
Kaiser, N., Squires, G. \& Broadhurst, T.\ 1995, ApJ 449, 460.

\bibitem{}{}{}
Kneib, J.-P., Mellier, Y., Fort, B. \& Mathez, G.\ 1993, A\&A 273,
370. 

\bibitem{}{}{}
Kneib, J.-P., Mathez, G., Fort, B., Mellier, Y., Soucail, G. \&
Longaretti, P.-Y.\ 1994, A\&A 286, 701.

\bibitem{}{}{}
Kneib, J.-P.\ et al.\ 1995, A\&A 303, 27.

\bibitem{}{}{}
Kneib, J.-P., Ellis, R.S., Smail, I., Couch, W.J. \& Sharples, R.M.\
1996, ApJ 471, 643. 

\bibitem{}{}{}
Link, R. \& Pierce, M.J.\ 1998, astro-ph/9802207.

\bibitem{}{}{}
Luppino, G. \& Kaiser, N.\ 1997a, ApJ 475, 20.

\bibitem{}{}{}
Luppino, G. \& Kaiser, N.\ 1997b, talk given at the ``Workshop on Weak
and cluster 
gravitational lensing'', Ringberg Castle, Jan.\ 1997.

\bibitem{}{}{}
Miralda-Escud\'e, J.\ 1991, ApJ 380, 1.

\bibitem{}{}{}
Natarajan, P. \& Kneib, J.-P. 1997, MNRAS 287, 833.

\bibitem{}{}{}
Natarajan, P., Kneib, J.-P., Smail, I. \& Ellis, R.S.\ 1998, ApJ 499,
600. 

\bibitem{}{}{}
Olsen, L.F.\ et al.\ 1998, astro-ph/9803338.

\bibitem{}{}{}
Postman, M. et al.\ 1996, AJ 111, 615.

\bibitem{}{}{}
Schneider, P.\ 1996, MNRAS, 283, 837.

\bibitem{}{}{}
Schneider, P.\ 1998, ApJ 498, 43.

\bibitem{}{}{}
Schneider, P., Ehlers, J. \& Falco, E.E.\ 1992, {\it Gravitational
lenses}, Springer: New York.

\bibitem{}{}{}
Schneider, P. \& Rix, H.-W.\ 1997, ApJ 474, 25.

\bibitem{}{}{}
Schneider, P., van Waerbeke, L., Mellier, Y., Jain, B., Seitz, S. \&
Fort, B.\ 1998a, A\&A 333, 767.

\bibitem{}{}{}
Schneider, P., van Waerbeke, L., Jain, B. \& Kruse, G.\ 1998b, MNRAS
296, 873.

\bibitem{}{}{}
Seitz, S.\ 1998, talk given at the IAP Meeting May 1998, Paris.

\bibitem{}{}{}
Seitz, C., Kneib, J.-P., Schneider, P. \& Seitz, S.\ 1996, A\&A 314, 707.

\bibitem{}{}{}
Seitz, S., Saglia, R.P., Bender, R., Hopp, U., Belloni, P. \& Ziegler,
B.\ 1998, MNRAS, in press (also astro-ph/9706023).

\bibitem{}{}{}
Seitz, C. \& Schneider, P.\ 1995, A\&A 297, 287.

\bibitem{}{}{}
Seitz, S. \& Schneider, P.\ 1996, A\&A 305, 383.

\bibitem{}{}{}
Seitz, S., Schneider, P. \& Bartelmann, M.\ 1998, A\&A submitted, aslo
astro-ph/9803038.

\bibitem{}{}{}
Seljak, U.\ 1997, astro-ph/9711124

\bibitem{}{}{}
Smail, I., Ellis, R.S., Fitchett, M.J. \& Edge, A.C.\ 1995, MNRAS 273, 277.

\bibitem{}{}{}
Squires, G. et al.\ 1996a, ApJ 461, 572.

\bibitem{}{}{}
Squires, G., Kaiser, N., Fahlman, G., Babul, A. \& Woods, D.\ 1996b,
ApJ 469, 73. 

\bibitem{}{}{}
Tyson, J.A.\ 1988, AJ 96, 1.

\bibitem{}{}{}
Tyson, J.A., Valdes, F. \& Wenk, R.A. 1990, ApJ 349, L1.

\bibitem{}{}{}
Van Waerbeke, L.\ 1998, A\&A, in press (also astro-ph/9710244).

\bibitem{}{}{}
Van Waerbeke, L., Mellier, Y., Schneider, P., Fort, B. \& Mathez, G.\
1997, A\&A 317, 303.

\bibitem{}{}{}
Wambsganss, J., Cen, R. \& Ostriker, J.P.\ 1998, ApJ 494, 29.


\end{thebibliography}
\end{document}